\documentclass{aa}  

    \usepackage{graphicx}
    \usepackage{txfonts}
    \usepackage[colorlinks = true, allcolors=blue]{hyperref}
    \usepackage{stfloats}
    \usepackage{booktabs}
    \usepackage{orcidlink}
    \usepackage{sidecap}
    \usepackage{pdflscape}
    \newcommand\arcdeg{\mbox{$^\circ$}}%
    \newcommand{\Msun}{\hbox{M$_\odot$}}
    
\begin{document}

       \title{UOCS. XIV. Uncovering extremely low mass white dwarfs and blue lurkers in NGC 752\thanks{Full version of Table~\ref{tab:SED_results} is available in electronic form at the CDS via anonymous ftp to \url{cdsarc.cds.unistra.fr} (130.79.128.5) or via \url{https://cdsarc.cds.unistra.fr/cgi-bin/qcat?J/A+A/}}}
       \titlerunning{Blue lurkers in NGC 752}
       
       \author{Vikrant V. Jadhav\orcidlink{0000-0002-8672-3300}\inst{1,2},
              Annapurni Subramaniam\orcidlink{0000-0003-4612-620X}\inst{3},
              Ram Sagar\orcidlink{0000-0003-4973-4745}\inst{3}}
    
       \institute{
            Helmholtz-Institut für Strahlen- und Kernphysik, Universität Bonn, Nussallee 14-16, D-53115 Bonn, Germany\\
            \email{vjadhav@uni-bonn.de}
            \and
            Inter-University Centre for Astronomy and Astrophysics (IUCAA), Post Bag 4, Ganeshkhind, Pune 411007, India
            \and
            Indian Institute of Astrophysics, Koramangala II Block, Bangalore-560034, India
            \email{purni@iiap.res.in}
            }
    
       \date{Received May 18, 2024; accepted June 6, 2024}
    
     
      \abstract
       {
       Evolutionary pathways of binary systems are vastly different from single stellar evolution, and thus, there is a need to quantify their frequency and diversity. Open clusters are the best test-bed to unveil the secrets of binary populations due to their coeval nature. And the availability of multi-wavelength data in recent years has been critical in characterising the binary population.
       }
       { 
       NGC 752 is a solar metallicity, intermediate-age open cluster located at 460 pc.
       In this work, we aim to identify the optically subluminous white dwarfs in NGC 752 and identify the illusive blue lurkers by association.
       }
       {We used multiwavelength photometry from \textit{Astrosat}/UVIT, \textit{swift}/UVOT, \textit{Gaia} DR3 and other archival surveys to analyse the colour-magnitude diagrams and spectral energy distributions of 37 cluster members.
       }
       {We detected eight white dwarfs as companions to cluster members. Four of the systems are main sequence stars with extremely low mass white dwarfs as their companions. Two are these main sequence stars are also fast rotators. 
       }
       {
       The presence of low mass white dwarfs and high rotation signals a past mass transfer, and we classified the four main sequence stars as blue lurkers. The binary fraction in NGC 752 was estimated to be 50--70\%, and it shows that the contribution of optically undetected stars is crucial in quantifying the present-day binary fraction.
       }
    
    
       \maketitle
    %

\section{Introduction} \label{sec:introduction}

A good fraction of stars in the Galaxy are part of binary or multiple systems. The evolution of stars in such systems can be different due to interactions with a close companion. Due to the varied nature of binary orbital parameters, predicting the nature of the interaction and its final product is not always possible. We are studying open clusters (OCs) to identify optically sub-luminous hot companions in binary systems and detect the signs of mass transfer in these systems.

In this work, we studied a nearby intermediate-age OC NGC 752 ($\alpha_{J2000} = 01^h 58^m, \delta_{J2000} = +37\arcdeg 52\arcmin$). Table~\ref{tab:ngc752_info} gives the basic parameters of the cluster.
It has been studied substantially using imaging \citep{Ebbighausen1939ApJ....89..431E, Eggen1963ApJ...138..356E, Crawford1970AJ.....75..946C, Arribas1990Ap&SS.169...49A, Platais1992BICDS..40....5P, Twarog2015AJ....150..134T} and spectroscopy \citep{Hobbs1986ApJ...309L..17H, Pilachowski1989PASP..101..991P, Hobbs1992AJ....104..669H, Bocek2020MNRAS.491..544B, Boesgaard2022ApJ...927..118B}. The cluster was also studied using X-rays \citep{Belloni1996A&A...305..806B, Giardino2008A&A...490..113G}.
NGC 752 has a moderate binary fraction of $\approx$40\% \citep{Jadhav2021AJ....162..264J}.
\citet{Maderak2013AJ....146..143M} noted an main sequence (MS) turn-off mass of 1.82 \Msun\ and reported O overabundance in cool dwarfs in the cluster.

\begin{table}
    \centering
    \begin{tabular}{ccr}
    \toprule
        Parameter & Literature values & This work\\ \hline
        Age & 1.45$\pm$0.05 [3], 1.34$\pm$0.06 [4],  & 1.58\\
        (Gyr) & 1.41 [5], 1.52 [7] & {$\pm$0.11}\\
        Distance & 474.2 [3], 438$\pm$8 [4], 448 [7], & 461\\
        (pc) & 483$\pm$15 [8], 443.8 [9] & $\pm$13\\
        E(B$-$V) & 0.048$\pm$0.009 [1], 0.034$\pm$0.004 [3], & 0.0435\\
        (mag) & 0.05 [5], 0.035 [7], 0.024 [8]  & {$\pm$0.0050}\\
        {[}Fe/H{]} & 0.16$\pm$0.09 [1], -0.063$\pm$0.013 [2],   & 0.0\\
        (dex) & $-$0.071$\pm$0.014 [3], 0.0 [5] & {$\pm$0.1}\\ 
    \toprule
    \end{tabular}
    \caption{Basic parameters of NGC 752. 
    \\References:
    [1] \citet{Barta2011BaltA..20...27B}, 
    [2] \citet{Maderak2013AJ....146..143M}, 
    [3] \citet{Twarog2015AJ....150..134T}, 
    [4] \citet{Agueros2018ApJ...862...33A}, 
    [5] \citet{Siegel2019AJ....158...35S},
    [6] \citet{Lum2019ApJ...878...99L},
    [7] \citet{Bocek2020MNRAS.491..544B}, 
    [8] \citet{Cantat2020A&A...640A...1C}, 
    [9] \citet{Agarwal2021MNRAS.502.2582A}
    }
    \label{tab:ngc752_info}
\end{table} 

Recently, \citet{Buckner2018RNAAS...2..151B} reported discovery of MS+white dwarf (WD) system 
in NGC 752 using \textit{Gaia} photometry and panchromatic spectral energy distributions (SEDs).
\citet{Milone2019AJ....158...82M} studied an eclipsing SB2 system, DS Andromedae, 
and estimated dynamical properties of the components.
\citet{Bhattacharya2021MNRAS.505.1607B} found tidal tails around the cluster spanning 35 pc. They also found that the cluster has lost 92.5--98.5\% mass due to stellar evolution and tidal interactions. 
\citet{Sandquist2023AJ....165....6S} analysed two eclipsing binaries in the turn-off of NGC 752 and postulated non-standard evolution for both binaries. 
The cluster is well separated in proper motion space and has a well-established list of members. \citet{Cantat2020A&A...640A...1C} listed 223 members brighter than 18 Gmag while \citet{Agueros2018ApJ...862...33A} listed 258 members in F0--M4 spectral range. For further analysis, we use the members from \citet{Cantat2020A&A...640A...1C} due to the better precision of the \textit{Gaia} data.

\begin{figure*}
    \centering
    \includegraphics[width=0.95\textwidth]{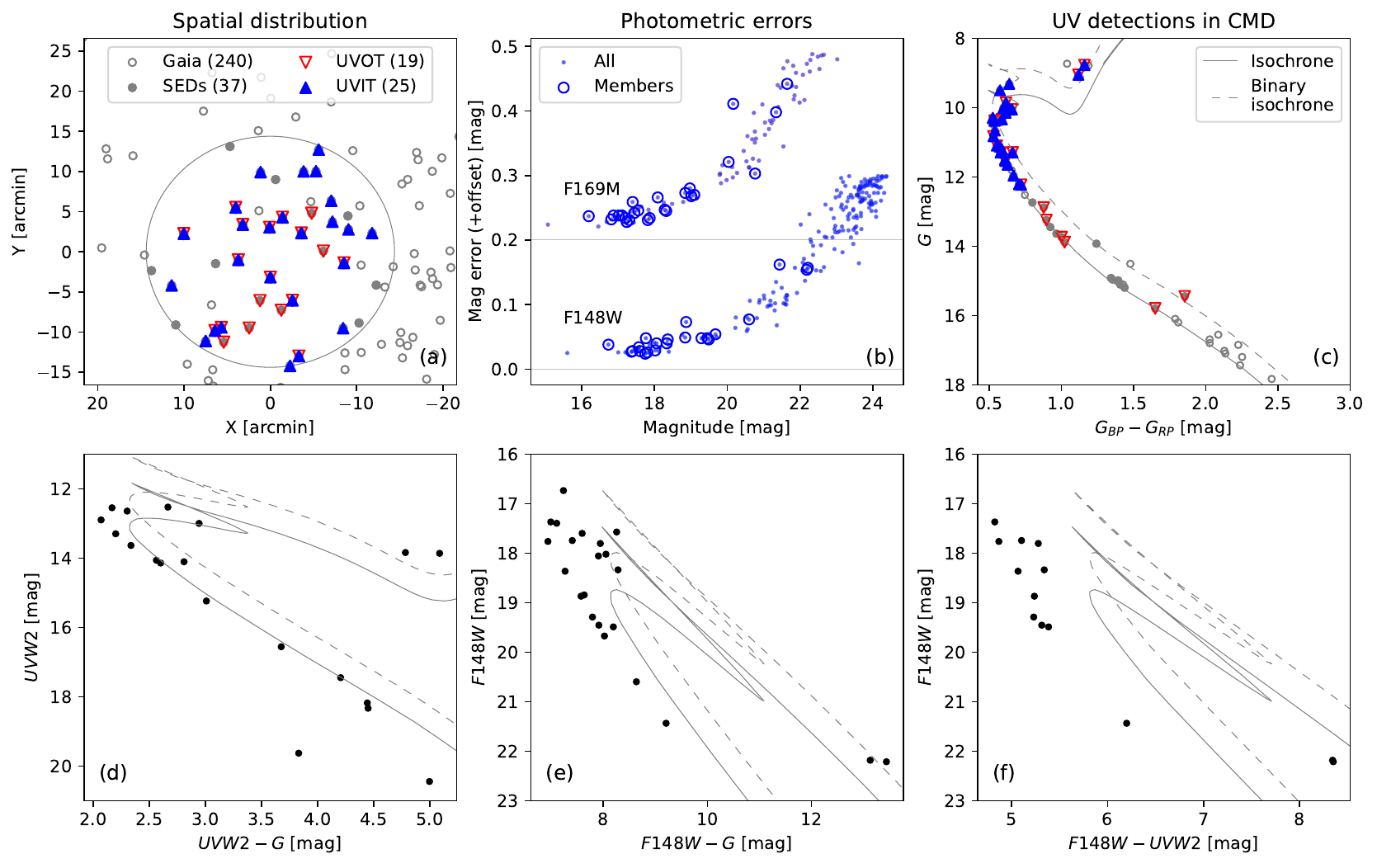}
    \caption{(a) Spatial distribution of NGC 752 members and UV detections. The \textit{Gaia}-based cluster members (grey circles), UVOT detections (red triangles), UVIT detections (blue-filled triangles) and the stars brighter than 15.8 Gmag analysed with SED fitting (grey-filled circles) are shown. The large grey circle denotes the UVOT field of view. (b) Photometric errors in UVIT/F148W and UVIT/F169M photometry. Cluster members are shown as blue circles, while all UV detections are shown with dots. The magnitude errors in the F169M filter are offset by 0.2 mag for clarity. (c) \textit{Gaia} CMD of the cluster members and their UV detections. The symbols are the same as panel (a). (d) UVOT-\textit{Gaia} CMD. (e) UVIT-\textit{Gaia} CMD. (f) UVIT-UVOT CMD. The reddened PARSEC isochrone (grey curve) and binary isochrone (grey dashed curve) are shown for reference in panels (c)--(f).}
    \label{fig:spatial_dist}
\end{figure*}

In an OC, the mass of young WDs is predefined based on its MS turn-off mass. For NGC 752, the young WDs should have a mass of $\approx$ 0.5 \Msun\ \citep{Maderak2013AJ....146..143M, Cummings2018ApJ...866...21C}. However, there is evidence of detecting $\approx$ 0.2 \Msun young WDs in other OCs such as M67 \citep{Landsman1997ApJ...481L..93L, Jadhav2019ApJ...886...13J}. Evolution of such extremely low mass WDs (ELMs) is not possible through single stellar evolution within the Hubble time. Hence, they are products of binary interaction as mass donors \citep{Marsh1995MNRAS.275..828M}. The ELMs stay bright and hot relatively longer than higher mass WDs due to their thick atmospheres and lower initial temperature. Hence, it is much more common to detect ELMs when the companion has evolved (into a WD or a neutron star) and become optically sub-luminous \citep{Brown2010ApJ...723.1072B}. Detecting such optically dominant ELMs is also quite efficient \citep{Pelisoli2019MNRAS.488.2892P}. In contrast, detecting ELMs in the presence of an optically bright companion is much more challenging. As these young WDs are hot but compact, their optical flux is multiple magnitudes lower than the MS-like acceptor present in close proximity. To identify these unresolved binaries with different temperatures, a multi-wavelength SED can be used \citep{Jadhav2022arXiv220703780J}. The detection of an MS+ELM system can be used to confirm the system's mass transfer history. Complementarily, OCs are known to host post mass transfer systems such as blue stragglers \citep{Sandage1953AJ.....58...61S} and blue lurkers \citep{Leiner2019ApJ...881...47L}. 
{Blue stragglers are the stars more massive and bluer than the MS stars formed via mass transfer or collisions \citep{McCrea1964MNRAS.128..147M, Hills1976ApL....17...87H}. Blue lurkers are MS stars, with similar mass transfer history as the blue stragglers, which are identified based on their faster rotation \citep{Leiner2019ApJ...881...47L} or other mass transfer signatures \citep{Jadhav2019ApJ...886...13J}.}
The identification and frequency of blue stragglers in OCs and globular clusters has been well established \citep{Knigge2009Natur.457..288K, Jadhav2021MNRAS.507.1699J}. However, identifying blue lurkers is difficult due to their unremarkable position in the colour-magnitude diagram (CMD). Presently, only a handful of clusters have blue lurker candidates \citep{Jadhav2019ApJ...886...13J, Nine2023ApJ...944..145N, Dattatrey2023MNRAS.523L..58D} and the sample is highly incomplete due to the illusive and transient signatures of mass transfer.

In this work, we used multi-wavelength photometry of NGC 752 to detect ELM candidates in the cluster and increase the sample of known blue lurkers by association. The paper is organised as follows: Sect.~\ref{sec:data} present the data and analysis, we present and discuss the results in Sect.~\ref{sec:results} and summarise in Sect.~\ref{sec:conclusion}.

\section{Observations and Analysis} \label{sec:data}

\subsection{Data}

\begin{table}
    \centering
    \begin{tabular}{ccccc}
    \toprule
        Filter & $\lambda_{pivot}$ & Exp. time & No. of sources & Members \\
         & [\AA] & [s] & & \\ \hline
        F148W & 1481 & 4747 & 199 & 22\\
        F169M & 1609 & 2569 & 92 & 25 \\ \hline
        UVW1 & 2581 & 798 & 338 & 18 \\
        UVM2 & 2246 & 1032 & 301 & 18 \\
        UVW2 & 2055 & 1076 & 314 & 19 \\
    \bottomrule
    \end{tabular}
    \caption{Log of UVIT and UVOT observations.}
    \label{tab:UVIT_data}
\end{table}

\begin{table*}
    \centering
\begin{tabular}{lccc cccc cc}
\toprule
Name & Platais ID$^\dagger$ & RAdeg & DEdeg & T$_{eff\, A}$ & L$_{A}$ & MH$_{A}$ & T$_{eff\, B}$ & L$_{B}$ & M$_{WD}$\\
     &                      &  [\arcdeg]     & [\arcdeg]      &  [K]  &[L$_{\odot}$]&       &  [K]  &[mL$_{\odot}$]& \\
\midrule
star1 &     & 29.2828 & 37.8243 & 4000$_{-250}^{+250}$ & 0.08$_{-0.02}^{+0.02}$ & -0.5 & 14250$_{-500}^{+250}$ & 1.3$_{-0.3}^{+0.3}$ & 1.2\\
star14 & 955 & 29.4973 & 37.9149 & 6500$_{-250}^{+250}$ & 12.11$_{-2.48}^{+2.48}$ & -0.5 & 13750$_{-250}^{+250}$ & 38.7$_{-8.7}^{+8.5}$ & 0.2\\
star17 & 1000 & 29.5476 & 37.6592 & 6750$_{-250}^{+250}$ & 3.13$_{-0.64}^{+0.64}$ & 0.0 & 10750$_{-250}^{+250}$ & 120.3$_{-24.8}^{+24.7}$ & 0.16\\
star18 & 1089 & 29.6243 & 37.8603 & 5000$_{-250}^{+250}$ & 27.65$_{-5.67}^{+5.67}$ & 0.0 & 9500$_{-250}^{+250}$ & 18.4$_{-3.9}^{+3.8}$ & 0.18\\
star30 & 689 & 29.2633 & 37.929 & 6500$_{-250}^{+250}$ & 2.31$_{-0.48}^{+0.47}$ & 0.0 & 14250$_{-250}^{+250}$ & 8.7$_{-2.2}^{+2.2}$ & 0.5\\
star33 & 641 & 29.2211 & 37.8692 & 6500$_{-250}^{+250}$ & 9.23$_{-1.89}^{+1.89}$ & -0.5 & 15250$_{-250}^{+250}$ & 20.8$_{-4.5}^{+4.5}$ & 0.3\\
star35 & 580 & 29.1635 & 37.8614 & 7000$_{-250}^{+250}$ & 8.1$_{-1.66}^{+1.66}$ & 0.0 & 15250$_{-250}^{+250}$ & 41.5$_{-8.6}^{+8.6}$ & 0.2\\
star43 & 1117 & 29.6538 & 37.7529 & 6750$_{-250}^{+250}$ & 16.8$_{-3.44}^{+3.44}$ & -0.5 & 13750$_{-250}^{+250}$ & 95.0$_{-21.6}^{+21.7}$ & 0.19\\ \hline
star3 & 772 & 29.3364 & 37.8619 & 6500$_{-250}^{+250}$ & 10.0$_{-2.05}^{+2.05}$ & 0.0 & 12500$_{-250}^{+250}$ & 55.3$_{-11.4}^{+12.8}$ & 0.19\\
star6 & 824 & 29.3827 & 37.8945 & 6750$_{-250}^{+250}$ & 2.67$_{-0.55}^{+0.55}$ & 0.0 & 12500$_{-250}^{+250}$ & 18.3$_{-3.8}^{+3.8}$ & 0.2\\
star8 & 867 & 29.4123 & 37.77 & 5000$_{-250}^{+250}$ & 36.27$_{-7.44}^{+7.44}$ & 0.0 & 12750$_{-500}^{+1750}$ & 1.1$_{-0.5}^{+0.6}$ & 1.2\\
star9 & 868 & 29.4144 & 37.8737 & 6750$_{-250}^{+250}$ & 7.35$_{-1.52}^{+1.52}$ & -0.5 & 11250$_{-250}^{+250}$ & 234.0$_{-49.1}^{+48.0}$ & 0.17\\
star13 & 950 & 29.4908 & 37.8061 & 6500$_{-250}^{+250}$ & 3.23$_{-0.66}^{+0.66}$ & 0.0 & 11500$_{-250}^{+250}$ & 36.9$_{-7.6}^{+7.6}$ & 0.19\\
star16 & 988 & 29.5321 & 37.6658 & 6750$_{-250}^{+250}$ & 4.94$_{-1.01}^{+1.01}$ & -0.5 & 11750$_{-250}^{+250}$ & 124.4$_{-27.3}^{+28.6}$ & 0.18\\
star40 & 890 & 29.4364 & 37.9884 & 6500$_{-250}^{+250}$ & 10.89$_{-2.23}^{+2.23}$ & -0.5 & 15000$_{-250}^{+250}$ & 21.3$_{-4.4}^{+4.4}$ & 0.3\\
\bottomrule
\end{tabular}
    \caption{SED parameters of NGC 752 members with significant UV excess. The top eight source have no X-ray detection. The bottom seven sources have X-ray emission hence the SED fitted parameters of the B component may be unreliable. $^\dagger$ The ID number from \citet{Platais1992BICDS..40....5P}. An extended version of the table, including UVIT photometry and SED fitting parameters, is available at the CDS.}
    \label{tab:SED_results}
\end{table*}

\begin{figure*}
    \centering
    \includegraphics[width=0.95\textwidth]{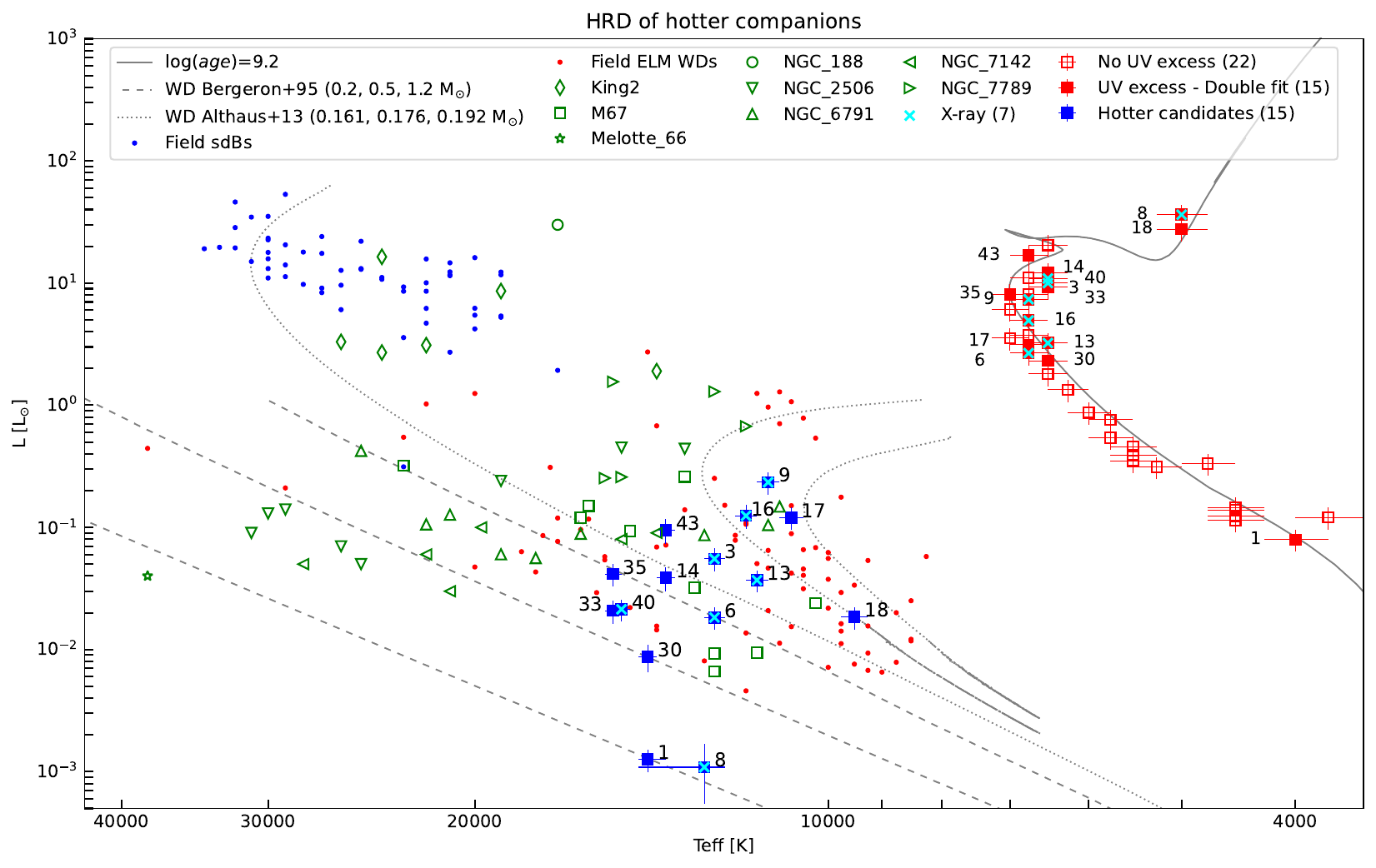}
    \caption{HRD positions of NGC 752 members and their companions. The NGC 752 members are indicated as follows: Members with UV excess and binary fit (red-filled squares) and hot companion candidates (blue-filled squares). Cyan crosses mark the X-ray-detected sources.
    Previously detected compact companions in OCs King 2 (green diamonds), M67 (green squares), Melotte 66 (green star), NGC 188 (green circles), NGC 2506 (green down-pointing triangles), NGC 6971 (green up-pointing triangles), NGC 7142 (green left-pointing triangles) and NGC 7789 (green right-pointing triangles) are also shown. 
    The cluster isochrone (grey curve), {\citet{Bergeron1995PASP..107.1047B} WD cooling curves (grey dashed curves), \citet{Althaus2013A&A...557A..19A} WD cooling curves (grey dotted curves)} and HRD positions of field sdBs (blue dots) and ELMs (red dots) are shown for reference.}
    \label{fig:HRD}
\end{figure*}

NGC 752 was observed with UltraViolet Imagin Telescope (UVIT) onboard \textit{AstroSat} on 2019-Dec-27 in two far-UV filters. The details of the observations are given in Table~\ref{tab:UVIT_data}. The calibration and instrumentation details of UVIT can be found in \citet{Tandon2017AJ....154..128T, Tandon2020AJ....159..158T} and \citet{Kumar2012SPIE.8443E..1NK} respectively. 
The UVIT images were processed using \textsc{ccdlab} to create science ready images \citep{Postma2017PASP..129k5002P, Postma2020PASP..132e4503P, Postma2021JApA...42...30P}. The PSF photometry was performed using \textsc{iraf} \citep{Tody1993ASPC...52..173T}. The UVIT photometric catalogue is available in the electronic version of the Table~\ref{tab:SED_results} at the CDS.

We used \citet{Cantat2020A&A...640A...1C} catalogue for cluster membership. The sources with $proba>0.5$ are considered members for further analysis.
Among the 212 UVIT-detected sources, 25 are cluster members. We also checked the Ultra-violet Optical Telescope (UVOT)/\textit{swift} catalogue and found 19 cluster members \citep{Siegel2019AJ....158...35S}\footnote{\url{archive.stsci.edu/hlsp/uvot-oc}}. The total number of cluster members with at least UVIT or UVOT detection was 31 sources.
Table~\ref{tab:UVIT_data} gives more information about the source detection in UV. 
The optically faintest UV member had a magnitude of $\sim$15.8 Gmag. Hence, we selected all 39 sources brighter than 15.8 Gmag within the same field of view for further SED analysis. This enabled us to study the UV properties of a G-band magnitude-limited sample of cluster members.

Fig.~\ref{fig:spatial_dist} (a) shows the spatial distribution of \textit{Gaia} DR2 members and UV detected members. Fig.~\ref{fig:spatial_dist} (b) shows the error distribution for UV images. Fig.~\ref{fig:spatial_dist} (c) shows the \textit{Gaia} CMD of the cluster, including UV-detected sources. Fig.~\ref{fig:spatial_dist} (d)--(f) show the UV-optical CMDs of NGC 752.  

We checked the source locations in aladin\footnote{\url{https://aladin.u-strasbg.fr/AladinLite/}} and \textit{Gaia} DR3 to check for crowding within 5\arcsec. Two stars  (star4, star27) out of the 39 were removed from further SED analysis due to the presence of close neighbours.
In addition to UVIT and UVOT photometry, we used \textsc{vosa} \citep{Bayo2008A&A...492..277B} to search the UV detected sources in photometric archives: 2MASS \citep{Skrutskie2006AJ....131.1163S}, AKARI/IRC \citep{Murakami2007PASJ...59S.369M, Onaka2007PASJ...59S.401O}, \textit{WISE} \citep{Wright2010AJ....140.1868W}, Pan-STARRS \citep{Chambers2016arXiv161205560C}, $Uvby\beta$ photoelectric photometric catalogue \citep{Hauck1998A&AS..129..431H}, \textit{Gaia} DR3 \citep{Gaia2022arXiv220800211G}.
The photometry was corrected for reddening using \citet{Fitzpatrick1999PASP..111...63F}, \citet{Indebetouw2005ApJ...619..931I} and \citet{Castelli2003IAUS..210P.A20C} extinction laws.
In addition to imaging data, 27 stars within the sample have APOGEE-DR17 spectroscopic data \citep{Majewski2017AJ....154...94M, Abdurrouf2022ApJS..259...35A}.

\subsection{Isochrone fitting}
We used PARSEC isochrones with [M/H] = 0.0 and a log(age) range of 9.05 to 9.20 (with steps of 0.03) to get the best-fitting isochrone. 
The fitting was done visually; the parameters are given in Table~\ref{tab:ngc752_info}. {The given errors are the grid sizes used in the fitting.}
The solar metallicity 1.58 Gyr PARSEC isochrone with the distance of 461 pc and E(B$-$V) of 0.0435 are plotted in Fig. \ref{fig:spatial_dist} and \ref{fig:HRD}.

\subsection{SED fitting}
We used a distance of $461\pm37$ pc to cover the mean distance from isochrone and the literature values in Table~\ref{tab:ngc752_info}. Similarly, the extinction of 
E(B$-$V) of $0.0435\pm0.0100$ ($\equiv$ A$_V$ of $0.135\pm0.030$) is used to deredden the stellar fluxes.

The 37 isolated sources\footnote{None of these sources are variable based on \textit{Gaia} DR3} were fitted with single SED composed of a Kurucz spectrum (suitable for MS and giant components; \citealt{Castelli2003IAUS..210P.A20C}). The parameter range of the Kurucz models was as follows: T$_{eff}$ $\in$ [3500, 9000], log$\,g$ $\in$ [3, 5], [M/H] $\in$ [-0.5, 0] and alpha = 0.
The SED fitting was performed using \textsc{Binary\_SED\_Fitting} v3.3.0\footnote{\url{https://github.com/jikrant3/Binary_SED_Fitting}} \citep{Jadhav2021JApA...42...89J}. 
\textsc{Binary\_SED\_Fitting} performs a $\chi^2$ minimising grid to find the component's parameters. A few data points were removed from the sources for a good fit. These non-fitted points are indicated in the respective SEDs in Fig.~\ref{fig:SED_results}. 

We found that 15 out of 37 sources show UV excess flux in at least two UV filters. The excess was identified using fractional residual (FR), $\Delta flux/flux_{obs} > 0.5$. 
These sources were further fitted with a second component composed of a Koester spectrum suitable for WD components (\citealt{Koester2010MmSAI..81..921K, Tremblay2009ApJ...696.1755T}).
The parameter range for Koester models were as follows: T$_{eff}$ $\in$ [7000, 80000] and log$\,g$ $\in$ [6.5, 9.5].
The resultant binary SED fits are shown in Fig.~\ref{fig:SED_results} and tabulated in Table~\ref{tab:SED_results}. An extended version of the table is available online.


\section{Results and Discussion} \label{sec:results}
Fig.~\ref{fig:spatial_dist} shows the spatial location of NGC 752 members. There are 37 isolated members within a common field of view and the UV detection limit. The optical (Fig.~\ref{fig:spatial_dist} c) and near UV CMD (Fig.~\ref{fig:spatial_dist} d) show that the cluster members follow the theoretical isochrones. However, some of the MS turn-off members in UV CMDs (Fig.~\ref{fig:spatial_dist} d--f) appear brighter than the isochrones. This is an indication of UV excess in these stars. In the UVIT and UVOT combined data, we found significant UV excess in 15 sources (40\%).

Fig.~\ref{fig:HRD} shows the Hertzsprung--Russell diagram (HRD) of the isolated cluster members brighter than 15.8 Gmag. The optically bright members all lie along the theoretical isochrone. 
{The HRD positions of field ELMs (sub-sample from \citealt{Brown2016ApJ...818..155B}) and field sdO/sdBs (sub-sample from \citealt{Geier2020A&A...635A.193G}) are shown for comparison whose parameters were derived using \textsc{vosa} \citep{Jadhav2023A&A...676A..47J}.}
The hotter components from binary SED fitting are also shown as blue squares. Their HRD positions are similar to the field ELMs. For reference, we have also plotted HRD positions of known hotter companions in other OCs: King 2 \citep{Jadhav2021JApA...42...89J}, M67 \citep{Sindhu2019ApJ...882...43S, Jadhav2019ApJ...886...13J, Subramaniam2020JApA...41...45S, Pandey2021MNRAS.507.2373P}, Melotte 66 \citep{Rao2022MNRAS.516.2444R}, NGC 188 \citep{Subramaniam2016ApJ...833L..27S}, NGC 2506 \citep{Panthi2022MNRAS.516.5318P}, NGC 6791 \citep{Jadhav2023A&A...676A..47J}, NGC 7142 \citep{Panthi2024MNRAS.527.8325P} and NGC 7789 \citep{Vaidya2022MNRAS.511.2274V}.

We compared the HRD position of hotter companions to WD cooling curves to estimate the photometric mass. 
{We used \citet{Bergeron1995PASP..107.1047B} for WDs more massive than 0.2 \Msun\ and \citet{Althaus2013A&A...557A..19A} for less massive WDs.}
Of the 15 WD candidates, 10 have photometric masses of $\leq$0.2 \Msun. Seven of these WD candidates are also detected in X-rays \citep{Giardino2008A&A...490..113G}. The source of the X-ray could also contaminate the UV flux. Hence, we cannot be sure that the UV flux solely comes from a UV bright WD and cannot trust the resultant SED parameters. This still leaves eight WD candidates in NGC 752, five of which are ELM candidates: four MS+ELM (star14, star17, star35, and star43) and one giant+ELM (star18).
{In addition, the high mass WD with star1 suggests that it had a massive progenitor, likely a blue straggler.}

The APOGEE survey provides v\,sin$i$ measurements, which are indicators of the rotation velocity of the dwarf stars. Higher rotation has been linked to recent mass accretion and indicator of a blue lurker \citep{Leiner2019ApJ...881...47L}. 
Among the WD candidates, four stars (star17, star30, star33 and star35) have v\,sin$i$ measurements (with values of 15--96 km s$^{-1}$). The star17, star33 and star35 have low mass WD companions (0.2--0.3 \Msun), with star33 having the highest v\,sin$i$ of 96 km s$^{-1}$.
Overall, the MS+WD systems have higher v\,sin$i$ compared to stars without UV excess (v\,sin$i_{median}$ = 5.4 km s$^{-1}$). The enhanced v\,sin$i$ supports the recent mass transfer history required to form the low mass companions in star17, star33, and star35.

The binary fraction derived using unresolved binaries in the optical CMD of NGC 752 is 28--48\% \citep{Jadhav2021AJ....162..264J}. From the optical CMD (Fig.~\ref{fig:spatial_dist} a), we can see that only a few UV-detected sources lie near the binary isochrone. The rest lie on the MS, which means they will not be included in the binary population based on the optical CMD.
The eight MS/giant+WD systems lead to a 22\% binary fraction among the 37 analysed systems (all of which lie on the MS in the optical CMD). Including these MS+WD system, the binary fraction of the cluster becomes 50--70\%. 
{The current work is sensitive to systems where the FR is more than 0.5, equivalent to an excess flux of 0.44 mag in at least two filters.
To achieve this, the bluest MS turn-off star (star35) would require a WD companion brighter than 19.2 mag in F148W. The UV magnitude-limit also constrains the amount of time such WD can be detected for a given mass: $<$1 Gyr for 0.19 \Msun\ \citep{Althaus2013A&A...557A..19A}, $<$100 Myr for 0.5 \Msun, and $<$270 Myr for 1.2 \Msun\ \citep{Bergeron1995PASP..107.1047B}.
Thus, a typical CO core hydrogen atmosphere WD (0.5 \Msun) in NGC 752 will be visible for only 100 Myr.
Comparatively, a He core WD cools down slower than the CO core WDs, thus leading to a higher detection rate.}
Overall, this demonstrates that the binary fraction estimates limited to optical analysis can lead to underestimating the present-day binary fraction. 


\section{Conclusions and summary} \label{sec:conclusion}
We analysed a magnitude limited sample of 37 members of NGC 752 (35 MS and two giants) using UVIT, UVOT, \textit{Gaia} and other archival data. 
\begin{itemize}
    \item The SED analysis showed that the cluster hosts at least eight WDs hidden in binary systems. Five WDs are ELMs and companions to four MS and one giant star. 
    \item There are four MS+ELM systems, two of which have higher rotation (v\,sin$i$), which is also a signature of recent mass transfer. Based on the ELM companion and high rotation, we classify these four sources as blue lurkers ($>$11\% of the MS population). Thus, NGC 752 is the third OC confirmed to contain blue lurkers after M67 and NGC 6791. 
        Six other MS stars with X-ray detection could also harbour an ELM companion. However, more analysis is needed to confirm their presence. 
    \item The binary fraction of MS+WD systems is 20\% (7/35). The binary fraction of NGC 752, accounting for the WD companions, is 50--70\% (22\% more than the binary fraction based on unresolved binaries in optical CMDs). A similar increase in the binary fraction of other clusters is also expected.
\end{itemize}

\begin{acknowledgements}
    VJ thanks the Alexander von Humboldt Foundation for their support.
    \textit{UVIT} project is a result of the collaboration between IIA, Bengaluru, IUCAA, Pune, TIFR, Mumbai, several centres of ISRO, and CSA. This publication makes use of {\sc VOSA}, developed under the Spanish Virtual Observatory project. 
    This work has made use of data from the European Space Agency (ESA) mission {\it Gaia} (\url{https://www.cosmos.esa.int/gaia}), processed by the {\it Gaia} Data Processing and Analysis Consortium (DPAC, \url{https://www.cosmos.esa.int/web/gaia/dpac/consortium}). Funding for the DPAC has been provided by national institutions, in particular the institutions participating in the {\it Gaia} Multilateral Agreement.
\end{acknowledgements}
\bibliographystyle{aa} 
\bibliography{references}

\begin{appendix}
\onecolumn

\section{Supplementary table and figures}

\begin{figure}[!h]
    \centering
    \includegraphics[width=0.95\textwidth]{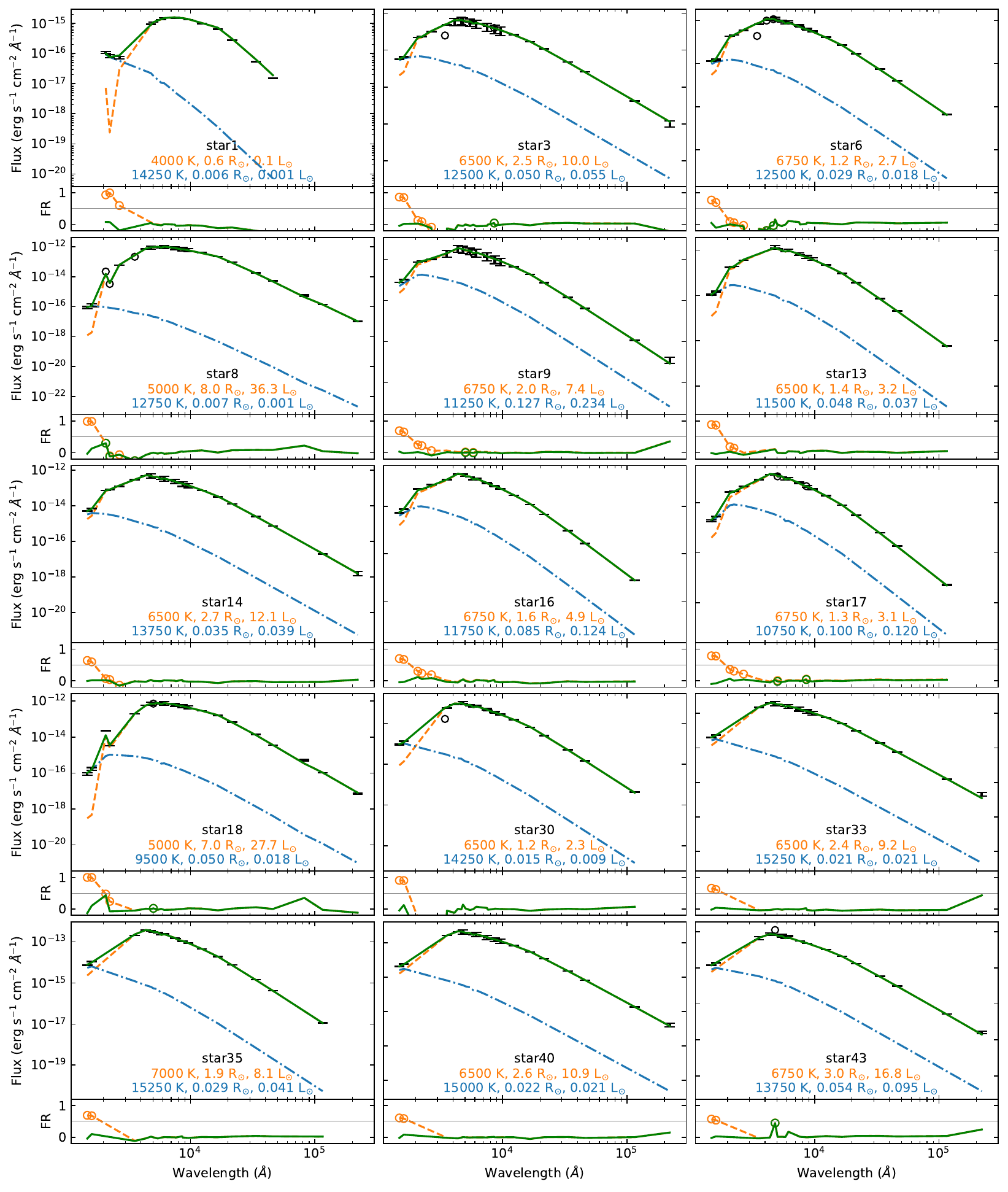}
    \caption{Binary SED fittings of NGC 752 members. The red, blue and green curves represent model SEDs of cooler component, hotter component and the binary system, respectively. The black error-bars and hollow circles show fitted, and non-fitted observed flux, respectively. The bottom panels in each SED shows the FR in each fit with colours similar to the SEDs.}
    \label{fig:SED_results}
\end{figure}

\end{appendix}

\end{document}